\documentclass[review]{elsarticle}

\usepackage{lineno,hyperref}
\usepackage{graphicx}
\usepackage{amsmath}

\modulolinenumbers[5]

\journal{Sensors and Actuators B: Chemical}

\begin{document}

\begin{frontmatter}

\title{Ultra high sensitive long period fiber grating based sensor for detection of adulterator in ethanol}

\author{Krishnendu Dandapat}
\address{Department of Physics, Indian Institute of Technology Kanpur, U.P., 208016, India}

\author{Indrajeet Kumar$^1$ and Saurabh Mani Tripathi$^{1,2}$}
\address{$^1$Department of Physics, Indian Institute of Technology Kanpur, U.P., 208016, India
$^2$Centre for Laser and Photonics, Indian Institute of Technology Kanpur, U.P., 208016, India}


\begin{abstract}
An ultra-sensitive sensor based on long-period fiber graings (LPFGs) has been fabricated for the detection of methanol and water content in ethanol. Our sensor is very compact in size, highly accurate and easy to fabricate making it very useful than the conventional surface plasmon resonance (SPR) sensor, which is bulky and expensive. We show that our sensor is capable to achieve an ultra sensitivity of 696.34 pm/ V$\%$ methanol and 655.3 pm/ V$\%$ water in presence of methanol and water in ethanol respectively. Our sensor is capable of detecting minimum  $1.5\times10^{-3}$ V$\%$ methanol and water in ethanol, which is far better than all the methods reported till now. 
\end{abstract}


\end{frontmatter}


\section{Introduction}

In daily life people use many liquors (alcoholis drinks) like wine, beer, brandy, whisky, ciders and vodka etc. Ethanol is an active ingredient of all alcoholic beverages. So, detection of adulterator in liquors (mainly in ethanol) is of prime importance for industrial research as well as human health. Also the determination of quality of alcohol is important in liquor industrial sectors. Easy and fast discrimation of the quality of liquor can help people to avoid consuming fake alcoholic drinks. Fake alcohol are often made of cheap industrial alcohol which can cause serious adverse effects on human body especially in developing country\cite{Pohanka1}. Commonly used ingredients to make fake alcohol drinks are methanol, isopropanol, automobile screen wash, nail polish remover as well as cleaning fluids. These fake alcohols are potentially very dangerous and can cause nausa, vomiting, abdominal pain, drowsiness, and dizziness. Methanol, ia an alcohol much like ethanol, also known as wood alcohol is the main substitute of ethanol used by illegal breweries. Methanol forms at low level during fermentation, is more dangerous than ethanol, can causes several illness like metabolic acidosis, neurologic sequelae, permanent blindness, liver damage, kidney failure and even death, when consumed\cite{Pohanka1}. Small amount of methanol is produced during fermentation, is fine to consume in commercial beer or wine because commercial producers have special methods to remove large amount of methanol from their products. Presence of large amount of methanol in alcoholic beverages is very harmful to human body. Most of the time, home brewers do not have specialized methods to remove large amount of methanol, can poison people.

Nowadays, researcher became interested to the search for greener and efficient options for energy production because of limited source of fossil fuels in nature and pollution caused by them. Among all the other possibilities, ethanol biofuel has shown suitable ehergy efficiency for combustion engines (70 $\%$ of gasoline power per gallon), low greenhouse gas emission rates. Ethanol can be produced from corn, sugarcane, and molasess, which is a byproduct in the process of making sugar. To use ethanol as a biofuel we need atleast 93 $\%$ pure ethanol. Mainly, ethanol is used as gasoline additive (approx. 10$\%$ vol.) in the United States, while in Brazil cars use gasoline additive, ethanol mixture (approx. 25$\%$ vol. of anhydrous ethanol fuel (AEF)) or simply ethanol (hydrated ethanol fuel (HEF)) as fuel\cite{Vello}. The main problem in ethanol as fuel, is the the addition of water as adulterator above the standard level. The presence of water in ethanol fuel can cause engine malfunction\cite{Queiroz}. The adulterator is difficult to figure out becuse of the high ethanol/water miscibility and colorless nature of the mixture. The contamination of ethanol is a major issue during distillation process which can cause major problem for fuel, chemical, pharmaceutical and beverage industries. In this present situation, we need a fast, efficient, accurate and low-cost sensor for the detection of adulterator in ethanol.

Over the years, many techniques to measure the quality of alcoholic beverages and alcohol concentration in liquors, such as distillation method\cite{Nykanen2}, high-pressure liquid chromatography\cite{Lazarus3}, gas chromatography\cite{Goldberg4} and wet chemical tritations\cite{Vallejos5} have been developed and reported. 

LPFGs are periodic refractive index variations inscribed inside the core region of optical fiber. LPFGs help to couple part of the optical field from fiber core region (radius $\sim$ 4.1 $\mu m$) known as core mode to the cladding region (radius $\sim$ 62.5 $\mu m$ ) known as the cladding mode of the optical fiber. The particular wavelength at which the coupling occurs between core mode and cladding mode is known as the resonance wavelength ($\lambda_R$) and can be expressed as\cite{Ruchi},
\begin{equation}
\lambda_R=\Lambda\times\bigg(n^{co}_{eff}-n^{cl}_{eff}+\frac{\kappa_{co-co}-\kappa_{cl-cl}}{k_0}\bigg)
\end{equation}
where $\kappa_{co-co}$ and $\kappa_{cl-cl}$ are the self-coupling coefficients of the core mode and cladding mode respectively; $\Lambda$ is the grating period; and $n^{co}_{eff}$ and $n^{cl}_{eff}$ are the effective refractive indices of the core mode and cladding mode respectively. Usually in optical fiber, core mode optical field propagates with most of the input power, is cofined in the core region, does not take part in sensing application. Whereas, cladding mode optical field carries very less power, spreads in the ambient region, very useful in sensing application. Periodic refractive index variations couple power from core mode to cladding mode, making LPFGs suitable for sensing application. A small portion of the cladding mode, known as evanescent field, propagates outside of the optical fiber, interacting with the external perturbations. Any variations in the refractive index of the ambient region will change $n^{cl}_{eff}$ and thus, will shift $\lambda_R$. The LPFG sensing principle primarily relies on measuring the shift in $\lambda_R$, due to change in the ambient refractive index. It is well known that higher the cladding mode order, larger would be the corresponding evanescent field in the region outside of the optical fiber, higher would be the sensitivity of the sensor.

There have been many attempts to increase the sensitivity of LPFG based sensor by coupling power to higher order cladding modes leading to the phenomenon of dual resonance\cite{Shu1}. In a dual-resonance LPG, power couples to same cladding mode at two distinct wavelengths. The two resonant wavelength shift towards opposite direction with any change in external perturbaton. This feature has huge apllication in sensing because it actually doubles the sensitivity of the LPG based sensor \cite{Shu1}\cite{Shu2}. If the grating period is close to the turn around point (TAP), sensitivity will be maximum \cite{Shu3}. 

In the present study, we report a very compact in size, highly accurate, easy to fabricate, and inexpensive long-period fiber gratings based sensor to detect methanol and water content in ethanol. We fabricate single resonance LPFG using KrF excimer laser and make it dual-resonace to maximize the sensitivity of the sensor by reducing the cladding radius. In order to avoid temperature and strain cross sensitivity we maintain a constant tension along the sensor and also constant temperature ($\sim 23^\circ C$) throught the experiment process. Our sensor shows an ultra sensitivity of 696.34 pm/ V$\%$ methanol and 655.3 pm/ V$\%$ water in presence of methanol and water in ethanol respectively. The reason of the resonance wavelength shift is the change of refractive index of the solution for different methanol and water content in ethanol. We also show that using our sensor we can measure minimum $1.5\times10^{-3}$ V$\%$ methanol and water in ethanol. 

\section{Materials and methods}
\subsection{Fabrication of the sensor}
We fabricated several LPFGs in hydrogen loaded (150 bar, for 15 days) telecommunication-grade single mode optical fiber (Corning, SMF-28e$^{TM}$) using a chromium amplitude mask ($\Lambda=215\mu m$) and high power KrF excimer laser (Lumonics Lasers, Pulse Master-840) emitting at 248 nm, at a pulse repetition rate of 100 Hz, pulse duration of 12 ns and peak pulse energy of 10 mJ \cite{Kd}. The LPFGs were then thermally annealed at 150 $^{\circ}$C for 3 hours to release excess hydrogen and stabilize the grating's optical properties. Finally the cladding region was partially etched in 10\% HF for $\sim$ 10 mins to tune into dual resonance region. 

\subsection{Materials}
Ethanol (99.9\% pure) was purchased from Changshu Yangyuan Chemical (China) and used without any further modification. Methanol (99.8\% pure) was purchased from Loba Chemical (India) and used in pure form for making different samples of ethanol-methanol solution. Disstilled water was purchased from Sigma-Aldrich (USA) for making different samples of ethanol-water solution and methanol-water solution. 

\subsection{Sample preparation}
Ethanol-methanol solution with different volume percentage of methanol in ethanol, were prepared by adding suitable volume of methanol in ethanol. After adding desired volume percentage of methanol in ethanol, the solution was stirred rigoriously for 10 mins. Samples of ethanol-water and methanol-water solution were prepared also using the above discussed method. The prepared sample's total volume was 10 ml for ethanol-methanol, ethanol-water, and methanol-water solution. To measure the volume, we used a micropipette having a smallest division of 0.5 ml. The refractive indices of each samples was measured by using Abbe's refractometer.

\subsection{Experimental Set-up}
The schematic diagram of the proposed sensor is shown in Fig. 1, where we use dual-resonance long-period fiber gratings (DRLPFGs) as sensing element. In our experiment one end of sensor is connected to a supercontinuum source (LEUKOS, SM-30-450) and the other end is connected to optical spectrum analyzer (YOKOGAWA, AQ6370D). The DRLPFG is very sensitive to any change in the ambient refractive index. Due to a change in the refractive index in the sensing region there will be spectral shift which is monitored by the optical spectrum analyzer.
\begin{figure}[h]
	\includegraphics[width=8 cm]{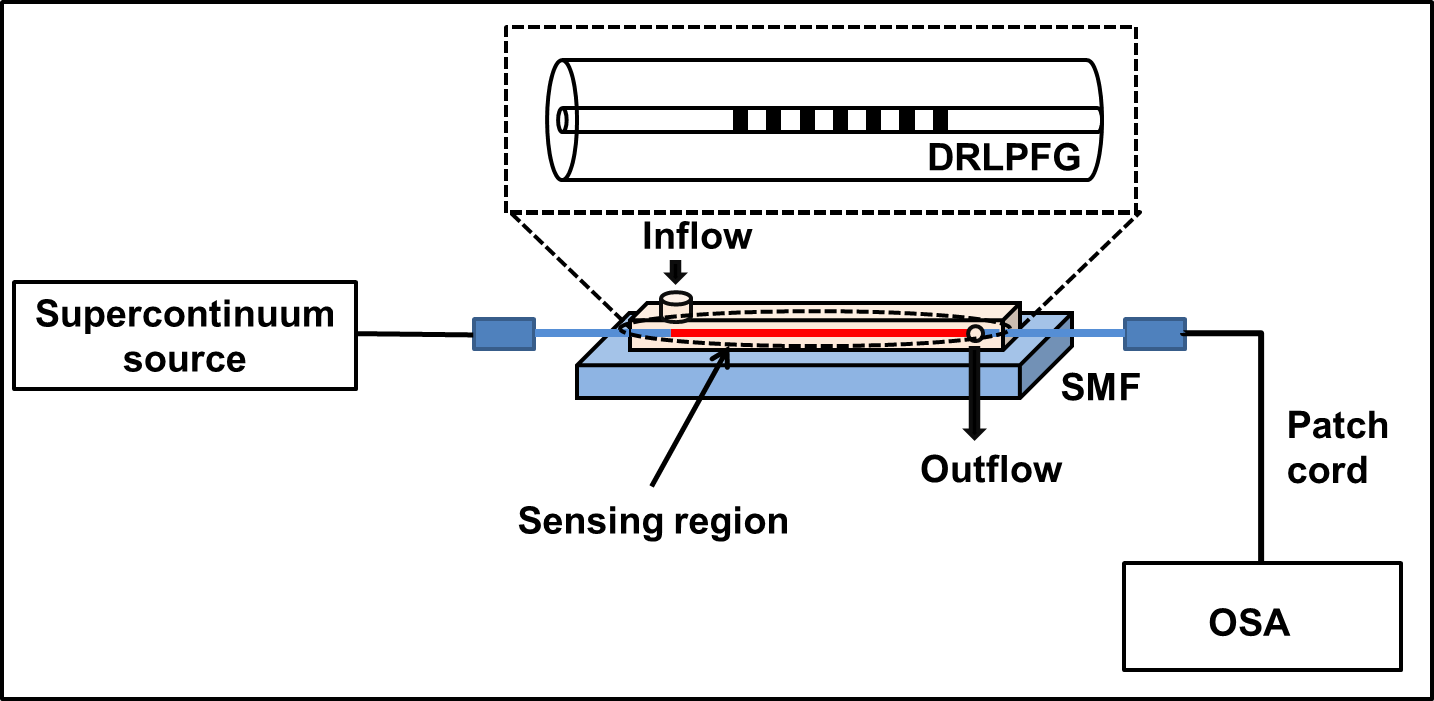}
	\caption{\textit{ Schematic diagram of the experimental set-up for the detection of adulterator in ethanol.}}
\end{figure} 
The presence of macrobends along the DRLPFG region is the primary cause of measurement errors in the DRLPFG based sensors. In order to avoid them we maintained a constant tension along the DRLPFG throughout the experiments by attaching the fiber near one end of the DRLPFG and applying a fixed force near the other end of it. To avoid any temperature cross sensitivity we maintain constant temperature ($\sim$ 23$^{\circ}$C) throughout our experiment process. During the measurement we used a movable stage and flow cell to confirm that our sensor is completely inside the samples and the spectral shift is only due to change in volume percentage of adulterator in ethanol. After each observation, the LPFG sensor surface was properly cleaned to remove contaminents. 

\section{Experimental results and discussion}

\subsection{Methanol volume percentage measurement in ethanol}
Samples of different volume percentage of methanol were prepared by disolving various volume of methanol in  ethanol. Transmission spectra for different volume percentage of methanol in methanol-ethanol solution were obtained to study the characteristics of the DRLPFG sensor. For a particular methanol volume percentage in methanol-ethanol solution, the DRLPFG sensor showed two unique dip in the transmission spectra. Depending on the volume percentage of methanol in methanol-ethanol solutin, the position of these two dip changes. The wavelength corresponding to the dips are called resonance wavelength ($\lambda_1 \& \lambda_2$). If the refractive index in the ambient region is high, the two dips will move away from each other and if refractive index is low will move towards each other. Transmission spectra of the DRLPFG sensor with different volume percentage of methanol in methanol-ethanol solution is shown in Fig. 2. 
\begin{figure}[h]
	\includegraphics[width=8 cm]{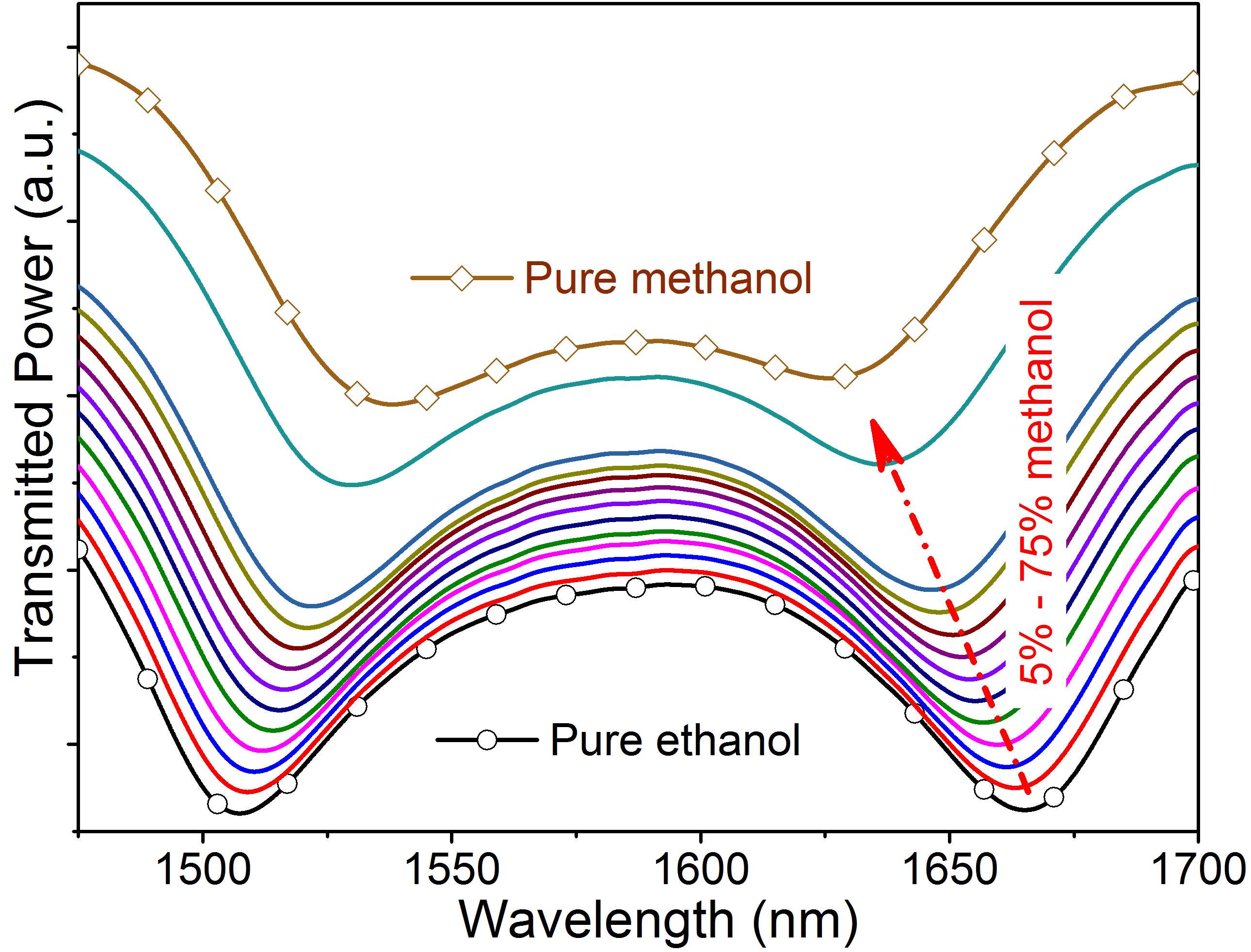}
	\caption{\textit{ Measured transmission spectra with different volume percentage of methanol in methanol-ethanol solution.}}
\end{figure}
Generally, the refractive index of a homogeneous mixture of two different solutions having different refractive index, is appeared to be between the individual refractive index of the two solutions. The individual refractive indices of pure ethanol and pure methanol is 1.36 and 1.327 respectively. The seperation of the two resonance dips doubles the sensitivity of the sensor. The seperation of the resonance dips with different volume percentage of methanol in methanol-ethanol solution is plotted in Fig. 3. The DRLPFG sensor shows a sensitivity of 696.34 pm/V$\%$ methanol in presence of methanol in methanol-ethanol solution, which is capable of detecting $1.5\times10^{-3}$ V$\%$ of harmful methanol in liquors using the resolution of 1 pm of the OSA used in our experiment. The spectral shift in the resonace wavelength with different V$\%$ of methanol in methanol-ethanol solution are plotted in Fig. 3.   
\begin{figure}[h]
	\includegraphics[width=8 cm]{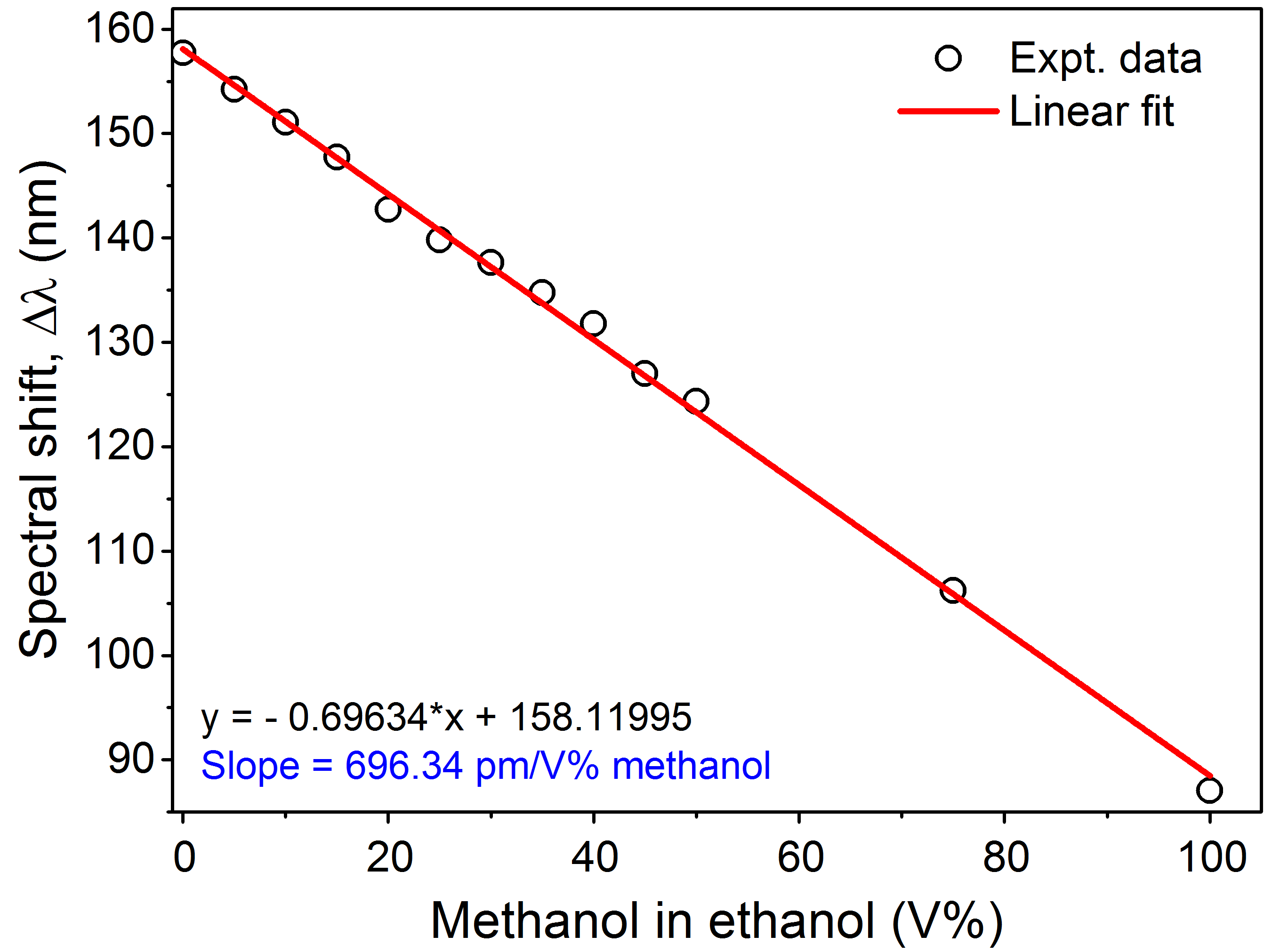}
	\caption{\textit{ Variation of wavelength shift with respect to different volume percentage of methanol in methanol-ethanol solution.}}
\end{figure} 
As we have discussed, if we vary the refractive index from high to low, the resonance dips will move towards each other. It is shown in Fig. 3 that if the volume percentage of methannol is increased in methanol-ethanol solution, the resonance dips are moving towards each other. The refractive index variation with different volume percentage of methanol in methanol-ethanol solution is shown in Fig. 4.
\begin{figure}[h]
	\includegraphics[width=8cm]{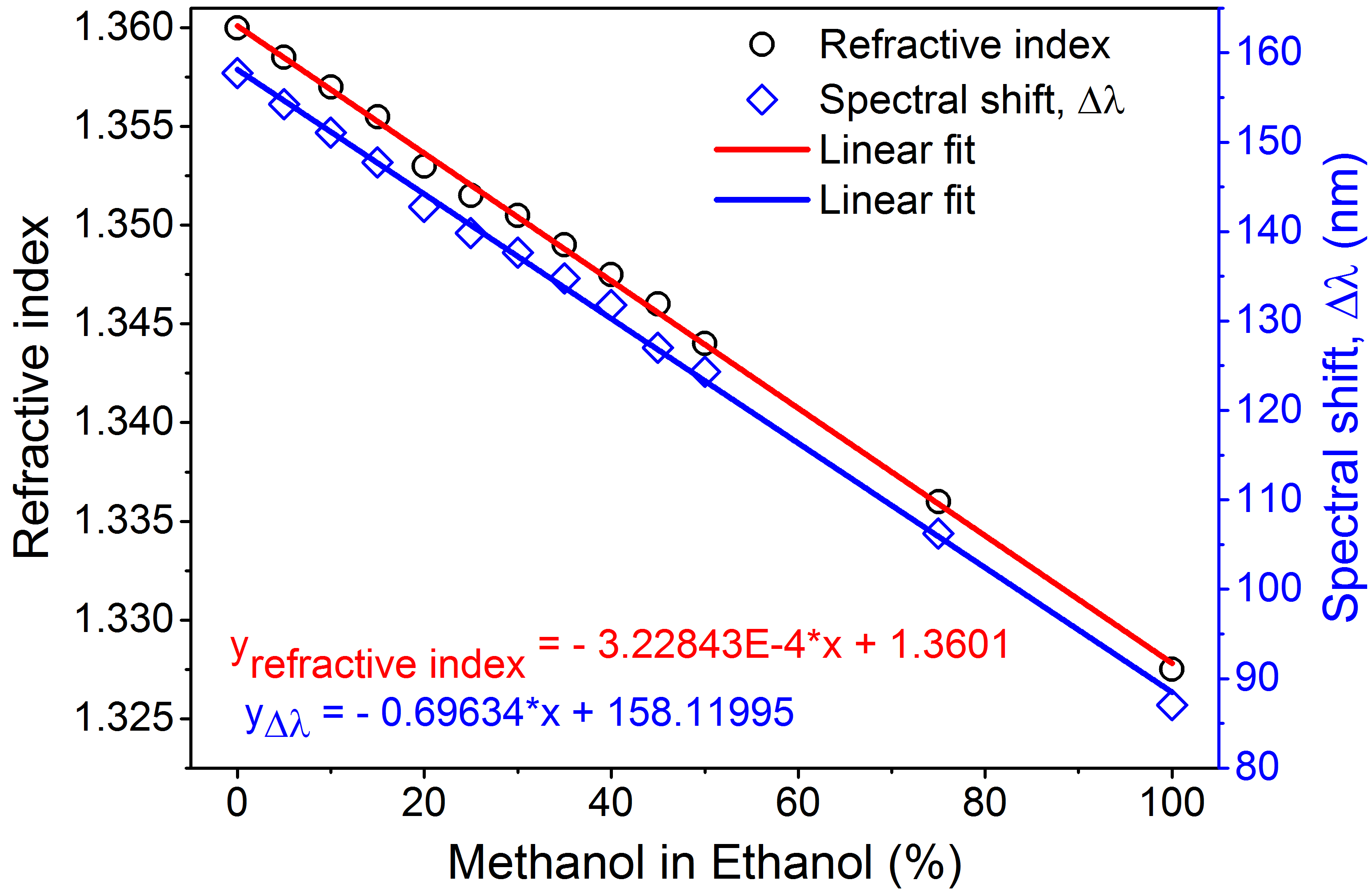}
	\caption{\textit{ Variation of the refractive index of the methanol-ethanol solution and spectral shift as a function of the volume percentage of methanol in methanol-ethanol solution.}}
\end{figure}
As shown in Fig. 4, the refractive index of the methanol-ethanol solution is decreasing in nature with increasing volume percenatge of metahnol. Becasue of this type of refractive index variation, resonace dips are moving towards each other.
\subsection{Water volume percentage measurement in ethanol}
Our sensor aslo capable of measuring water content in water-ethanol solution. Transmission spectra of the DRLPFG sensor with different volume percentage of water in water-ethanol solution is shown in Fig. 5.
\begin{figure}[h]
	\includegraphics[width=8cm]{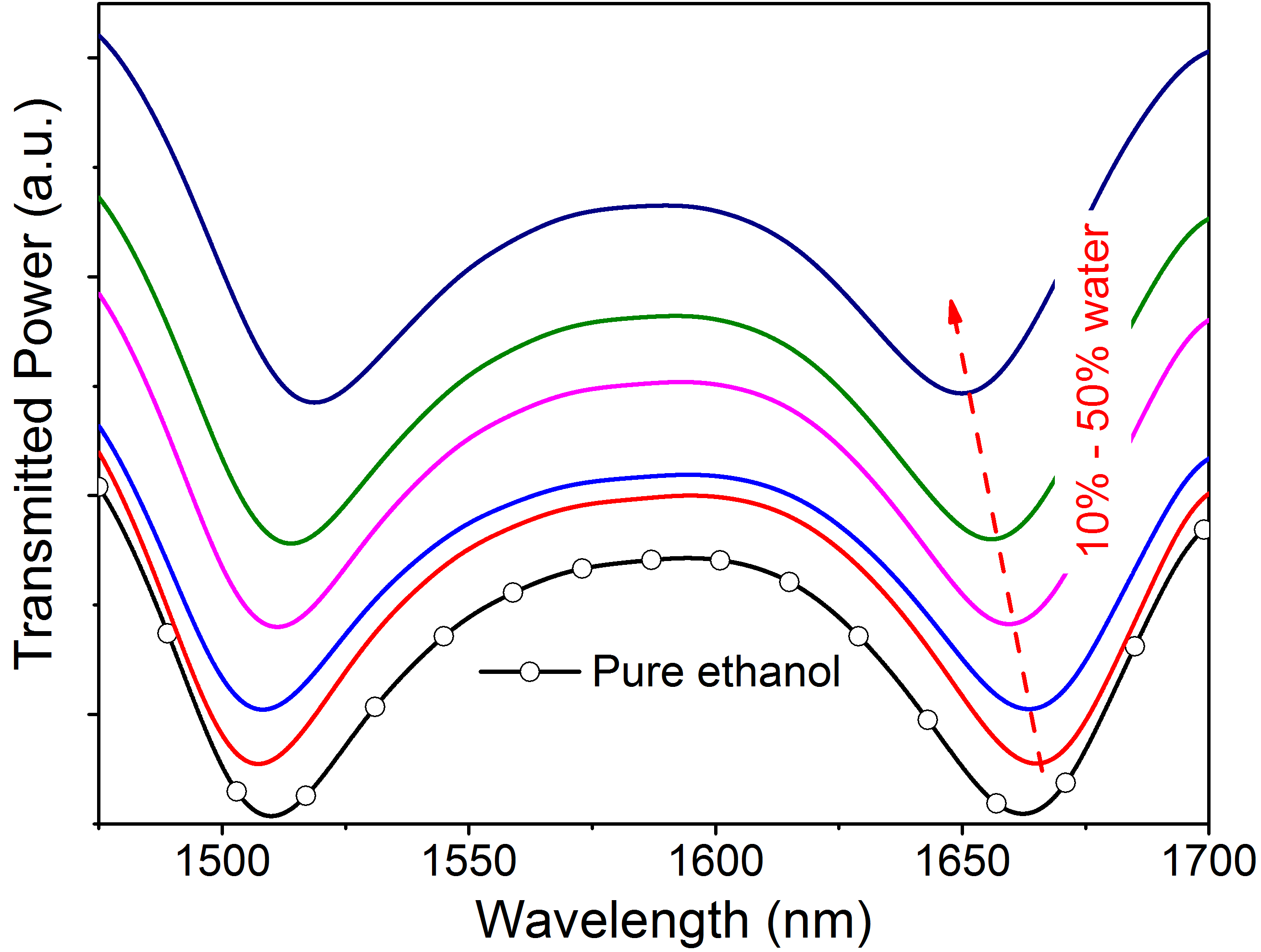}
	\caption{\textit{ Measured transmission spectra with different volume percentage of water in water-ethanol solution.}}
\end{figure}
As we have discussed earlier, the refractive index of a homogeneous mixture of two different solutions having different refractive index, should be between the individual refractive index of the two solutions. But from experiment, we observed that when the water content is small in ethanol, the spectral shift of the DRLPFG sensor is greater than pure ethanol and as we continued to increase the volume percentage of water in water-ethanol solution, the spectral shift decreases. Hence, we can conclude that when the water content is small,the refractive index of water-ethanol solution is greater than pure ethanol. Afetr 10 $\%$ volume percentage of water in the water-ethanol solution, the refractive index of the mixture started to decrease. The maximum refractive index is obtained at about 10 $\%$ volume percentage of water in ethanol. 
\begin{figure}[h]
	\includegraphics[width=8cm]{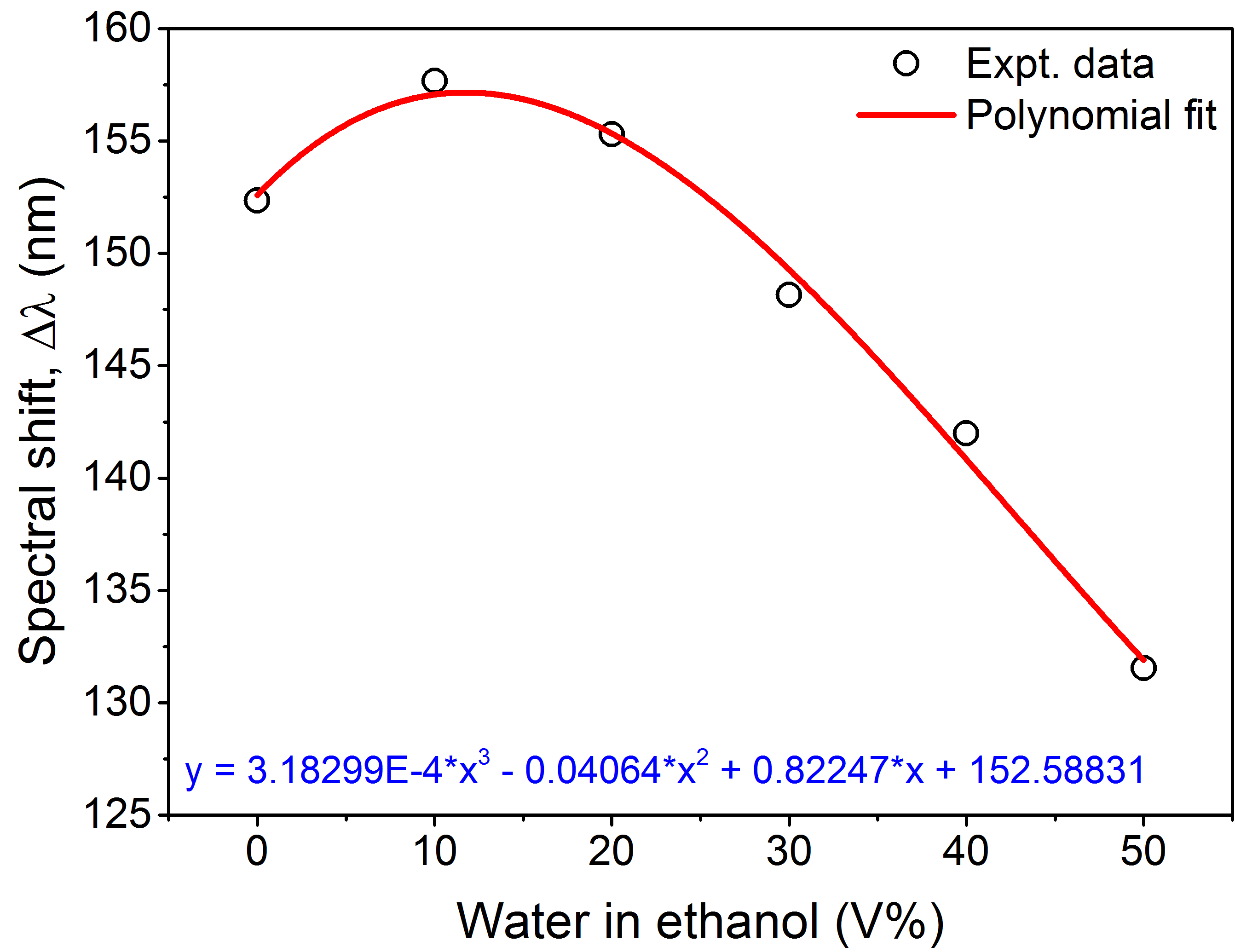}
	\caption{\textit{ Variation of wavelength shift with respect to different volume percentage of water in water-ethanol solution..}}
\end{figure}
From a macroscopic point of view, water and ethanol are completely miscible. But, from  a microscopic point of view, water and ethanol are non-miscible at presence of small water content ($<$ 10 V$\%$ of water). The reason behind the increase in refractive index due to small content of water is the formation of hydrogen bonds between water-ethanol \cite{Gereben}, \cite{Nishi}. Due to the formation of hydrogen bonds in water-ethanol solution, segregated clusters of either water or ethanol are formed. As a result, the formed clusters increase the density of the water-ethanol mixture and hence, the refractive index of the water-ethanol solution increases slightly \cite{Scott} \cite{BDGupta}. The spectral shift in the resonace wavelength with different V$\%$ of water in water-ethanol solution are plotted in Fig. 6.
\begin{figure}[h]
	\includegraphics[width=8cm]{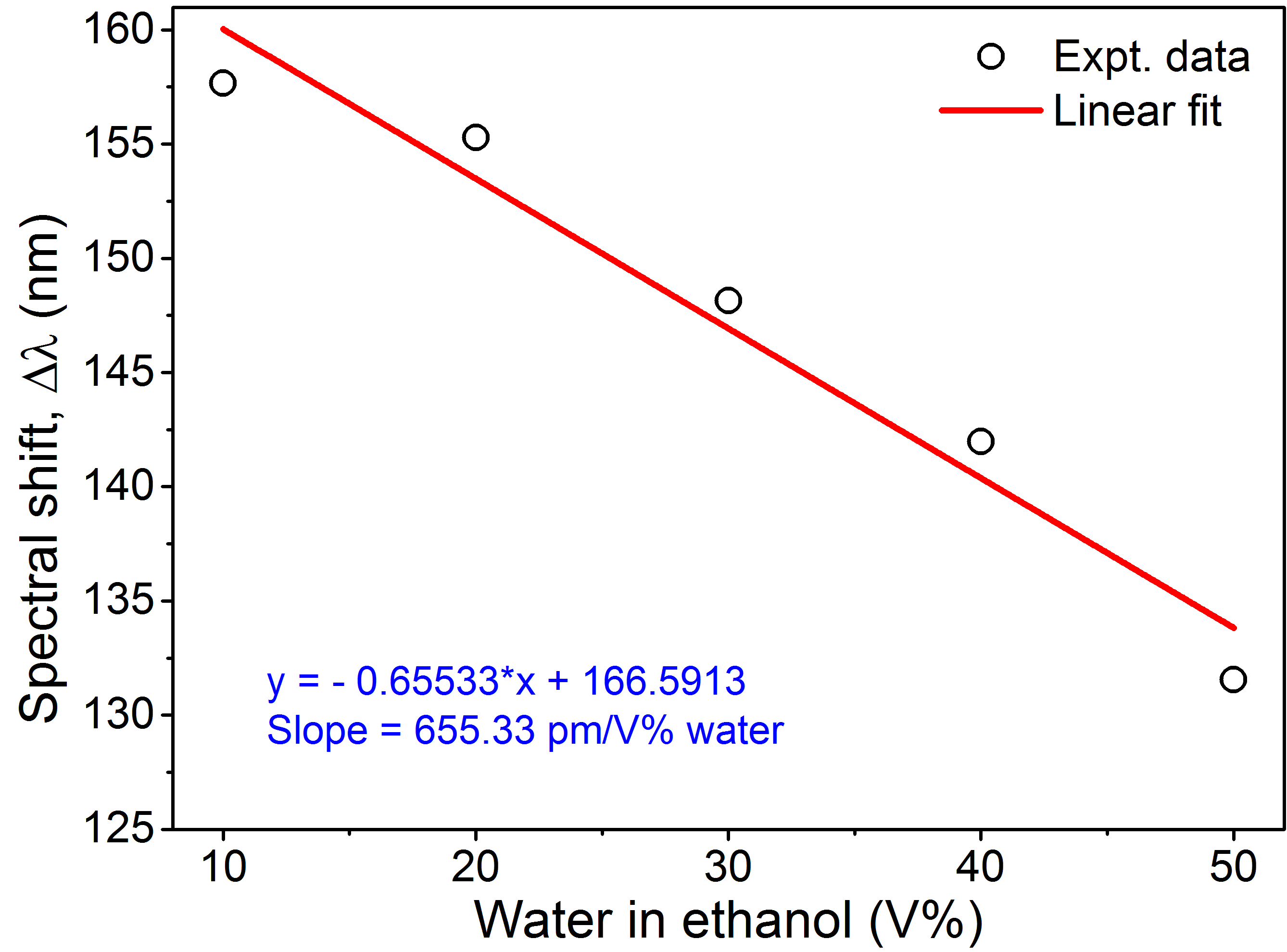}
	\caption{\textit{ Variation of spectral shift for 10-50 volume$\%$ of water in water-ethanol solution..}}
\end{figure}
The variation of wavelength shift as a function of 10-50 V$\%$ water in ethanol is plotted in Fig. 7, showing a linear relation. The sensitivity for the measurement of water content in ethanol is 655.33 pm/v$\%$ water. Using this sensor, we can measure a minimum of $1.5\times10^{-3}$ V$\%$ water in bio-fuel.

\section{Conclusions}
In this paper, we have demonstrated a very compact in size, highly accurate, easy to fabricate, and inexpensive long-period fiber gratings based sensor to detect methanol and water content in ethanol. We optimize the cladding diameter of the sensor to increase it's sensitivity. Our sensor shows a linear response for different V$\%$ of methanol in ethanol with a sensitivity of 696.34 pm/V$\%$ methanol. Similarly, for water content in range 10-50 V$\%$ in ethanol our sensor posses a linear response, showing an ultra-sensitivity of 655.33 pm/V$\%$ water. The sensitivity of our sensor is almost same and capable of detecting minimum $1.5\times10^{-3}$ V$\%$ of harmful methanol and water content in liquors. This sensor will be very useful for the measurement of adulterators in liquors and water content in ethanol which is mainly used as bio-fuel.

\section*{Acknowledgements}
The authors gratefully acknowledge financial support from the Ministry of Human Resource and Developement, India; Science and Engineering Research Board, Govt. of India.

\end{document}